# Optomechanical Generation of a photonic Bose-Einstein Condensate

Martin Weitz, Jan Klaers, and Frank Vewinger

*Institut für Angewandte Physik, Universität Bonn, Wegelerstr. 8, 53115 Bonn, Germany*

We propose to thermalize a low-dimensional photon gas and obtain photon Bose-Einstein condensation by optomechanical interactions in a microscopic optical cavity, with a single longitudinal mode and many transverse modes. The geometry of the short cavity is such that it provides a low-frequency cutoff at a photon energy far above the thermal energy, so that thermal emission of photons is suppressed and the photon number is conserved. While previous experiments on photon Bose-Einstein condensation have used dye molecules for photon gas thermalization, we here investigate thermalization owing to interactions with thermally fluctuating nanomechanical oscillators forming the cavity mirrors. In the quantum degenerate regime, the nanomechanical cavity converts broadband optical radiation into tuneable coherent radiation.

When a gas of particles is cooled, or its density is increased such that the associated de Broglie wavepackets spatially overlap, quantum statistical effects come into play. For bosonic particles with non-vanishing mass, Bose-Einstein condensation into a macroscopically occupied ground state then minimizes the free energy [1]. Bose-Einstein condensation has first been achieved in dilute atomic gases [2-3], and other experiments have reported (quasi-) equilibrium condensates of exciton-polaritons and magnons, respectively [4-7]. Blackbody radiation - the most ubiquitous Bose gas - does not show Bose-Einstein condensation, because the photon number is not conserved upon temperature variations (vanishing chemical potential), and at low temperatures photons disappear in the cavity walls instead of occupying the cavity ground state [1]. In recent work, we have observed Bose-Einstein condensation of a two-dimensional photon gas in a dye-solution filled optical microcavity [8]. The photon gas is here thermally coupled to the temperature of the dye-solution by multiple absorption-emission cycles in the dye, which conserves the average photon number [9]. The cavity mirrors provide both a non-vanishing effective photon mass and a confining potential.

The present proposal is motivated by recent advances in the field of optomechanics, the coupling of light and nanomechanical motion in optical cavities [10]. Thermal motion enhances the linewidth of optical radiation within an optical resonator [11,12]. The coupling strength of light and mechanical motion scales with the ratio of the zero-point mechanical motion and cavity length [10], and can be enhanced by use of thin optical cavities and small mirror masses. For wavelength-scale, nanomechanical mirrors usually direct thermal fluctuations of the mirror's center of mass positions will give the dominant contribution to the line broadening. Recent experiments based on microstructured nanomechanical oscillatiors have demonstrated mechanical oscillation frequencies in the GHz regime [13,14], approaching the transverse vibrational frequency of photons trapped in the microcavity experiment of [8]. While in the latter work absorption and emission processes were used for thermalization, we here show that the Doppler shifts (or Brillouin scattering) from thermally vibrating mirror oscillators can redistribute the optical frequency spectrum to a thermalized one, and under suitable conditions yield a macroscopic occupation of the ground state cavity mode. Moreover, we wish to draw attention to statistical aspects of light arising from such optomechanical interactions in optical multimode cavites.

Here we propose number conserving thermalization of a photon gas by means of optomechanical interactions in a microscopic optical cavity. Our setup uses a thin optical cavity, with a low-frequency cutoff far above the thermal energy and a manifold of transverse modes. Other than many current works in the field of optomechanics that operate under cryogenic conditions [10], we here assume the use of a room temperature environment. Thermal motion of the nanomechanical oscillators coupled to the photon gas leads to thermalization of the optical spectrum towards a Bose-Einstein distribution above the low-frequency cutoff. Above the BEC threshold phase space density, optomechanical interactions can redistribute a broadband optical spectrum (or more generally a nonequilibrium spectrum) into a Bose-Einstein condensed photon distribution, including a macroscopically occupied, coherent ground state mode.

In the proposed setup (Fig.1a), photons are confined in an optical microresonator with a vibrating mirror. The curved mirror cavity has many transverse modes, and the required optomechanical coupling between transverse modes suggests the use of a mechanically flexible or a segmented mirror, with a collection of microscopic springs attached to different mirror segments (indicated in Fig.1a), similar to multiple transducer mirror setups used in adaptive optics [15]. This allows the modification of the transverse optical wavevector required in the course of the thermalization, as a tilt of a mirror segment naturally occurs when the adjacent springs perform out of phase oscillations during their random thermal motion.

The mirror spacing is in the wavelength regime [8], and for simplicity assume that the mirror separation $d$ is near half an optical wavelength, fixing the longitudinal mode number to $q=1$. The tight longitudinal confinement introduces a low-frequency cutoff, the eigenfrequency of the $TEM_{00}$ transverse mode, at frequency $\omega_{cutoff} = 2\pi c / \lambda_{cutoff}$, with $\lambda_{cutoff} \cong d/2$. Figure 1b schematically shows the spectrum of transverse modes above the low-frequency cutoff, where the mirror curvature causes a harmonic trapping potential for the photons. Since the longitudinal mode number is fixed, the two remaining transverse modal degrees of freedom make the photon gas effectively two-dimensional. We are interested to obtain thermal equilibrium of the trapped photon gas, resulting in a distribution of width $\sim k_B T/\hbar$ above the cavity cutoff. In other words, significant population of the $TEM_{nm}$ modes with high transversal quantum numbers $n$ and $m$ will be reached at high temperature, while the distribution is limited to the lower transverse modes at low temperature.

The Doppler shift, which redistributes the light between different transversal modes, per cavity round-trip is relatively small, and many reflections within a high finesse cavity are required. The coupling is maximized when using mechanical oscillation frequencies $\omega_m$ of the micromechanical oscillators resonant to the transverse cavity mode spacing $\Omega$, and we correspondingly assume that $\omega_m \approx \Omega$. The mode spacing is $\Omega = c/\sqrt{dR}$ for a plano-concave mirror cavity, and typical values for $\Omega/2\pi$ are tens of *GHz* [8]. The vibrational frequency achieved in the optomechanical experiment of [14] was *3.7GHz*, which is still lower. Though we are aware that many polariton microcavity condensation experiments [4] operate even in the absence of confining potential (i.e. with $\Omega=0$), it seems advantageous to use some mirror curvature to both spatially confine the photon gas and enable a true equilibrium BEC at finite temperature in the two-dimensional system [16]. Mechanical vibrational frequencies of order of typical photon transverse trapping frequencies are hardly achievable with a single rigid mirror of sufficiently large diameter to capture the thermalized mode spectrum [17]. Instead, the use of a flexible mirror structure or a segmented geometry with many nanomechanical mirrors, as indicated in Fig.1a, seems a plausible solution. For infrared optical radiation with wavelength near $1.2\mu m$ or above, silicon-based nanostructures [14] can be used, while for the visible (or even the ultraviolet) spectral regime e.g. nanostructured glass materials will be required.

During thermalization it is essential to vary only transverse modal quantum numbers to leave the photon number constant (other than in a blackbody radiator). This condition is well achievable in the regime of $\hbar\omega_{cutoff} >> k_B T$. Thermal emission of photons is then suppressed by a factor of order $exp(-\hbar\omega_{cutoff}/k_B T)$, which is $\sim exp(-40)$ for $k_B T \cong 1/40eV$ (room temperature) and $\hbar\omega_{cutoff} \cong 1eV$ at $1.2\mu m$ cutoff wavelength.

Thermalization is mediated by photon scattering processes involving the absorption or emission of a phonon [18,19], as indicated in Fig.2a. Unlike recent works which discuss the strong coupling regime between mechanical motion and light [20], we assume that after many reflections, when thermalization is reached, the transverse decoherence from coupling to the thermal environment inhibits strong coupling effects to be relevant, as is also the case in [8]. Other than in the interparticle collisions of an atomic gas, the thermalization occurs with respect to an external reservoir, the movable mirror oscillators.

In the following, we aim at a derivation of the asymptotic distribution function of the photon gas using a detailed balance treatment [21]. We consider multiple reflections of light between one of the movable mirror segments and the fixed back mirror, as shown in Fig.2b, assuming a discretized version of a fully flexible mirror. For a given initial optical state $|1>$ we are interested in the probability for scattering into a final state |2> when the photon leaves this part of the cavity after a number N of reflections. Let |a> and |b> denote the initial and final states of the movable mirror respectively (who can e.g. differ in the number of phonons). In the case of a short phonon relaxation time the mirror segment quickly thermalizes and the occupation probabilites $P_a$ and $P_b$ of the mirror states follow a Boltzmann distribution, with $P_a / P_b = \exp\left(-\left(E_a - E_b\right)/k_B T\right)$. In the weakly coupled perturbative limit, the optomechanical coupling between different transverse cavity modes can be described by a rate equation model, which we derive using Fermi's golden rule $R = (2\pi/\hbar)\cdot\left|\langle f|V(b,a)|i\rangle\right|^2 \delta(E_f + E_b - E_i - E_a)$. Here the optomechanical interaction is written in the general form $V = \sum_{k,\ell} V_{k\ell}\, a_\ell a_k{}^\dagger$, where $V_{k\ell}(b,a) = V^*_{\ell k}(a,b)$. With initial and final states of the photon field $|i\rangle = |..,n_1,..,n_2,..\rangle$ and $|f\rangle = |..,n_1 - 1,..,n_2 + 1,..\rangle$ we obtain the total rates for a transfer 1,a → 2,b and 2,b → 1,a respectively

$$R_{1,a\to 2,b} = P_a C n_1 (n_2 + 1)\,, \tag{1}$$

$$R_{2,b\to 1,a} = P_b C n_2 (n_1 + 1)\,, \tag{2}$$

where $C$ is a constant containing the matrix elements, whose absolute values are not required in the following. We note that the same form of eqs.1-2 is well known from literature on stimulated (Brillouin) light scattering [22], a process equivalent to the Doppler shift picture of optomechanical interactions. In equilibrium due to detailed balance the transition rates are identical, from which follows

$$n_1/(n_1 + 1)\cdot P_a = n_2/(n_2 + 1)\cdot P_b\,. \tag{3}$$

If one assumes energy conservation, $E_1+E_a=E_2+E_b$ and that the equilibrium distribution functions $n_i$ of the photon modes depend only on the energy $E_i$, or $n_i/(n_i + 1) = f_i(E_i)$, we find $f_1(E_1)\exp\left(E_1/k_B T\right) = f_2(E_2)\exp\left(E_2/k_B T\right)$. In the general case $f_i(E_i)\exp\left(E_i/k_B T\right)$ has to be a fixed number, from which follows that $f_i(E_i) = e^{-(E_i - \mu_{ph})/k_B T}$ with $\mu_{ph}$ as a constant, which one identifies as the chemical potential, similar as in [21]. The distribution functions in thermal equilibrium thus can be written as $n_i = 1/(e^{(E_i - \mu_{ph})/k_B T} - 1)$, the usual Bose-Einstein distribution function [1].

The same result for the photon distribution is obtained if one takes the mirror states as fully quantum, i.e. when assuming a small phonon relaxation rate, so that bosonic stimulation will also apply for the states of the movable mirrors. The calculation then becomes formally equivalent to the detailed balance treatment given in [23] for the example of particle-particle (quantum) scattering.

The above derivation, for the sake of simplicity, assumes that the mirrors along with their mechanical suspension have sufficiently large thermal capacity that their temperature remains unchanged. For a finite heat capacity, depending on the wavelength of the incident pump radiation relatively to that of the emitted optical spectrum, the suspension will experience cooling or heating, respectively, by the optical radiation. Related optical cooling and heating effects are well known in optomechanics [10].

Let us next investigate the equilibrium distribution of the photon gas in the microresonator, following earlier work on the dye-filled system [8,9]. In the paraxial limit and for negligible interactions, the photon energy can be written as $E \cong m_{eff}c^2 + (\hbar k_r)^2 / 2m_{eff} + \frac{1}{2} m_{eff}\Omega^2 r^2$ , where we have defined an effective photon mass $m_{eff} = \hbar\omega_{cutoff} / c^2$ $(\equiv \hbar k_z(0)/c)$, and $k_r$ and $k_z$ denote transversal and longitudinal photon wavenumbers respectively. The system is formally equivalent to a two-dimensional, harmonically confined system of massive bosons, for which it is well known that a BEC exists at finite temperature [16]. If we account for the two-fold optical polarization degeneracy, the mode degeneracy at frequencies equal or above the cutoff, i.e. for $\omega \geq \omega_{cutoff}$ , is $g(\omega)=2((\omega-\omega_{cutoff})/\Omega+1)$ and we arrive at a frequency dependent photon number distribution $n(\omega) = g(\omega) / \left(e^{\hbar(\omega-\omega_{cutoff})-\mu)/k_B T} - 1\right)$, with an energy shifted definition of the chemical potential $\mu = \mu_{ph} - \hbar\omega_c$ to fulfill the usual convention of $\mu \rightarrow 0$ in the far quantum degenerate regime. At constant temperature, a BEC is expected once the photon number exceeds a critical particle number $N_c = \frac{\pi^2}{3}\left(\frac{k_B T}{\hbar\Omega}\right)^2$ . At room temperature, $T=300K$, and assuming a trap vibrational frequency $\Omega/2\pi = 10GHz$, we arrive at $N_c \cong 1.3 \cdot 10^6$, corresponding to a critical intracavity optical power of around $54W$, which is well feasible. Provided that thermalization is achieved, above the critical photon number a broadband optical spectrum is converted into a sharp condensate peak at the position of the cutoff on top of a broad thermal wing (see Fig.3a).

To estimate the thermalization time of the photon gas, we use a classical model based on optical sideband generation. For an optical beam of frequency $\omega$ passing through one of the partial cavities formed by one mirror segment and the back mirror, where $z(t) = z_0 \sin \omega_m t$ denotes the longitudinal displacement of a periodically vibrating mirror segment (measured at its center), the spatial movement will result in a periodic phase shift of the optical wave $\Delta\varphi(t) = 2\pi N \cdot 2z(t)/\lambda$ after *N* reflections (the geometry is as in Fig.2b). The optical wave is phase modulated by a modulation index $\beta = 4\pi N z_0/\lambda$, and for $\beta << 1$ the relative intensity in the first sidebands is $(\beta/2)^2 \cong (2\pi z_0/\lambda)^2 N^2$. When treating the thermally vibrating mirror segment as an harmonic oscillator in thermal equilibrium, one finds $z_0 = \frac{1}{\omega_m}\sqrt{2k_B T/m}$. As different mirror segments of the cavity oscillate independently, we estimate an average displacement

$$z_{0,av} = \sqrt{\frac{2}{N_{seg}}} z_0 = \frac{2}{\omega_m}\sqrt{\frac{k_B T}{m N_{seg}}}, \tag{4}$$

where we have assumed that both sides of the cavity consist of vibrating mirror segments, and the number of segments per mirror in the optical modal profile over that displacement must be averaged is $N_{seg}$. With a mechanical vibrational frequency $\omega_m$ chosen to be resonant with the transversal modal spacing, i.e. $\omega_m \cong \Omega$, the transfer of one optical vibrational quantum occurs at around $N \cong \lambda/(z_{0,av} 2\pi)$ (which is the condition for the sidebands to become as large as the carrier).

The deflection of the optical beam after *N* reflections for the case of a single segment due to the tilt of this segment is of order $\Delta\alpha = N \cdot 2z_0 / D$, where *D* denotes the diameter of the mirror segment. The value must again be averaged over the number of segments in the optical mode profile, and we find an average deflection $\Delta\alpha_{av} = \sqrt{2/N_{seg}} \cdot \Delta\alpha \cong \lambda/(D \cdot \pi)$. For a segment diameter *D* of order of the wavelength $\lambda$, the momentum spread will thus reach the width of the angular distribution of a thermalized optical spectrum $\Delta\alpha_{th} = \sqrt{\langle k_r^2 \rangle} / k_z(0) = \sqrt{2k_B T / m_{eff} c^2} \cong 0.22$, so that momentum conservation can be fulfilled during the course of the thermalization, and the mirror motion efficiently couples the transverse optical modes.

The energy transfer reaches the thermal energy after a number of reflections

$$\begin{aligned} N_{total} &\cong \frac{k_B T}{\hbar\Omega} \cdot N \\ &\cong \frac{\lambda}{2h}\sqrt{m k_B T N_{seg}}\,, \end{aligned} \qquad (5)$$

which can serve as an estimate for the required number of reflections in the cavity determining the thermalization timescale. Assuming a mass of the nanomechanical oscillator per mirror segment of $m \simeq 10^{-16} kg$, corresponding to the oscillating masses of the subwavelength sized silicon based microchip structures used in [14], and room temperature conditions (*T=300K*) we obtain $N_{total} \cong \sqrt{N_{seg}} \cdot 4.8 \cdot 10^8 \cdot \lambda / \mu m$. For a two-dimensional photon gas geometry, the number of mirror segments in the optical mode will be of order *100x100*, and we arrive at $N_{total} \simeq 5.8 \cdot 10^{10}$ for $\lambda \simeq 1.2 \mu m$. This is beyond currently achieved cavity quality factors, and clearly demands for further progress. Further, such a setup requires the application of three-dimensional structuring techniques [24]. An alternative would be the use of a one-dimensional photon gas geometry, for which a planar cavity structure could be used. This seems easier to fabricate using lithographic structuring of silicon microchip substrates, a technique that is well scalable to a large number of elements. The number of mirror segments here reduces to roughly $N_{seg} \simeq 100$, and we correspondingly estimate $N_{total} \simeq 5.8 \cdot 10^9$. Note that in this one-dimensional geometry the trapping potential needs to be (at least somewhat) more confining than parabolic to allow for formation of a BEC at finite temperatures [16]. This will result in a non-equidistant spacing of transversal modes, so that the resonance condition with the vibrating mirror segments cannot be accurately fulfilled throughout the transverse mode spectrum. A possible geometry, as a generalization of silicon based coupled optical nanomechanical oscillator geometry of [14], is shown in Fig.3b. With such photonic structures in principle also two-dimensional photonic quantum gases can be implemented using a three-dimensional mechanical structure, where again the longitudinal modal quantum number is frozen.

Other than in the model for the dye-filled microcavity BEC system presented in [25], where grandcanonical particle number fluctuations can occur due to effective particle exchange with the dye molecules, in the here discussed ‘empty-cavity’ system the photon number in an idealized experimental situation is strictly conserved. Correspondingly, only energy (but no number) exchange with the reservoir occurs, and the photon gas will be well described by a canonical ensemble, resulting in an expected ground mode intensity correlation function $g^{(2)}(0)=1$ in the condensed state.

To increase the oscillation amplitudes compared to thermal and thus increase the coupling between modes, one may in principle apply an external acoustic drive of the microcavity structure, e.g. from a series of high-frequency transducers similar as used in acousto-optical modulators driven with quasi-random radiofrequency signals. Although this will not result in a thermal equilibrium distribution of optical modes, it may nevertheless allow for a macroscopic accumulation of the optical ground state mode. An alternative method to achieve thermalization of the photon gas in a microresonator by optomechanical means is the scattering of light from nanospheres placed within the resonator, whose positions undergo thermal fluctuations, the well known Brownian motion [26]. A resonant coupling to the transverse cavity mode spacing will require the use of high frequency (~GHz) Brownian fluctuations.

To conclude, we propose to achieve thermalization of a low-dimensional photon gas in a microcavity by optomechanical interactions with the thermal environment. The conversion of broadband optical radiation into a Bose-Einstein condensed state, with a coherent, macroscopically occupied optical ground mode, is predicted by mechanical reflections from thermally vibrating mirror surfaces. Technical benefits of such a coherent light source would include the absence of a required coincidence with an atomic or molecular resonance line. Along similar lines, the proposed scheme also relieves one from optical power limitations due to the damage threshold of an intracavity optical gain or thermalization medium. From a fundamental viewpoint, it is believed that optomechanical light sources add a new concept to optical field synthetisation, which relies on a redistribution of optical frequencies, rather than on elementary emission (and/or absorption) processes of radiation.

We acknowledge support from the DFG (We1748-17) and the ERC (INPEC).

**Literature:**

[1] See, e.g.: K. Huang, *Statistical Mechanics*, 2nd edn, (Wiley, New York, 1987).

[2] M. H. Anderson et al., Science **269**, 198 (1995); K. B. Davis et al, Phys. Rev. Lett. **75**, 3969 (1995).

[3] See, e.g.: K. Bongs and K. Sengstock, Rep. Prog. Phys. **67**, 907 (2004).

[4] J. Kasprzak et al., Nature **443**, 409 (2006).

[5] R. Balili et al., Science **316**, 1007 (2007).

[6] See, e.g.: H. Deng, H. Haug, and Y. Yamamoto, Rev. Mod. Phys. **82**, 1489 (2010).

[7] S. O. Demokritov et al., Nature **443**, 430 (2006).

[8] J. Klaers, J. Schmitt, F. Vewinger, and M. Weitz, Nature **468**, 545 (2010).

[9] J. Klaers, F. Vewinger, and M. Weitz, Nature Phys. **6**, 512 (2010).

[10] See, e.g.: T. J. Kippenberg and K. J. Vahala, Science **321**, 1172 (2008).

[11] See, e.g.: G. J. Milburn and M. J. Woolley, Acta Phys. Slov. **61**, 483 (2011).

[12] V.B. Braginsky, M.L. Gorodetsky, and S.P. Vyatchanin, Phys. Lett. A **264**, 1 (1999); K. Numata, A. Kemery, and J. Camp, Phys. Rev. Lett. **93**, 250602 (2004).

[13] A. D. O'Connell et al., Nature **464**, 697 (2010).

[14] J. Chan et al., Nature **478**, 89 (2011).

[15] See, e.g.: R. K. Tyson, *Principles of Adaptive Optics*, 3nd edn, (CRC Press, Boca Raton, 2011).

[16] V. Bagnato and D. Kleppner, Phys. Rev. A **44**, 7439 (1991); W. J. Mullin, J. Low Temp. Phys. **106**, 615 (1997).

[17] The rms radius $r_{rms}$ of the thermal mode spectrum of a harmonically trapped photon gas is readily found using: $\frac{1}{2} m_{eff} \Omega^2 r_{rms}^2 = k_B T$ . Typical values at room temperature are at around *100$\mu$m*.

[18] See, e.g.: A. Heidmann, C. R. Physique **12**, 797 (2011).

[19] D. W. Snoke and S. M. Girvin, arXiv:1209.5369.

[20] S. Gröblacher, K. Hammerer, M. R. Vanner, and M. Aspelmeyer, Nature **460**, 724 (2009).

[21] See e.g.: M. Toda, R. Kubo, and N. Saito, *Statistical Physics I: Equilibrium Statistical Mechanics* (Springer, Berlin, 1992).

[22] W. Kaiser and M. Maier, in *Laser Handbook*, F.T. Arecchi and E.O. Schulz-Dubois (eds.) (North-Holland, Amsterdam, 1972), Vol. 2; see also: R. Loudon, *The quantum theory of light, second edition* (Clarendon Press, Oxford, 1983).

[23] See Chapter 7.2 in: W. Ketterle, D.S. Durfee, and D.M. Stamper-Kurn, in *Bose-Einstein condensation in atomic gases, Proceedings of the International School of Physics "Enrico Fermi", Course CXL*, M. Inguscio, S. Stringari, and C.E. Wieman (eds.) (IOS Press, Amsterdam, 1999), p. 67; arXiv:9904034.

[24] I. Staude et al., Opt. Lett. **35**, 1094 (2010).

[25] J. Klaers et al., Phys. Rev. Lett. **108**, 169403 (2012).

[26] See e.g.: E. Nelson, *Dynamical Theories of Brownian Motion* (Princeton University Press, Princeton, 1967).

**Figure captions:**

Fig.1 (a): Scheme of an optical microcavity consisting of one segmented moving mirror and a fixed back mirror. One of the mirrors is curved with a radius of curvature *R*. (b) Density of states of the microcavity. The small mirror spacing imposes a low-frequency cutoff frequency $\hbar\omega_{cutoff}$, corresponding to the eigenenergy of the *q=1* longitudinal and $TEM_{00}$ transverse mode. The indicated energy levels at larger frequencies give the mode spectrum of higher transverse modes of the *q=1* longitudinal modal manifold.

Fig.2 (a): Diagrams representing the scattering of a phonon, at a mirror resonance frequency $\omega_m$, with a photon of frequency $\omega$. (b) Representation of the optical path of a photon experiencing multiple reflections between one mirror segment and the cavity back mirror. The two springs indicate that the mirrors besides a longitudinal movement also can tilt to couple the transverse modes.

Fig.3 (a): The figure indicates the redistribution of optical frequencies from an input broadband optical spectrum (left) to a Bose-Einstein condensed spectrum (right), with a macroscopically occupied coherent BEC peak at the position of the cavity low-frequency cutoff on top of a thermal wing after thermalization from optomechanical interactions in the microcavity. (b) Schematic realization of a photonic crystal microcavity for optomechanical thermalization of a low-dimensional photon gas. The longitudinal (corresponding to an optical propagation along the z-axis) modal quantum number is frozen, and for a planar mechanical structure (as visible in the drawing plane) the photon gas is one-dimensional (also the y-direction is frozen). A two-dimensional photon gas can be realized with a three-dimensional mechanical structure, for which the patterning along the direction of the y-axis proceeds in the same way as shown for the x-direction.

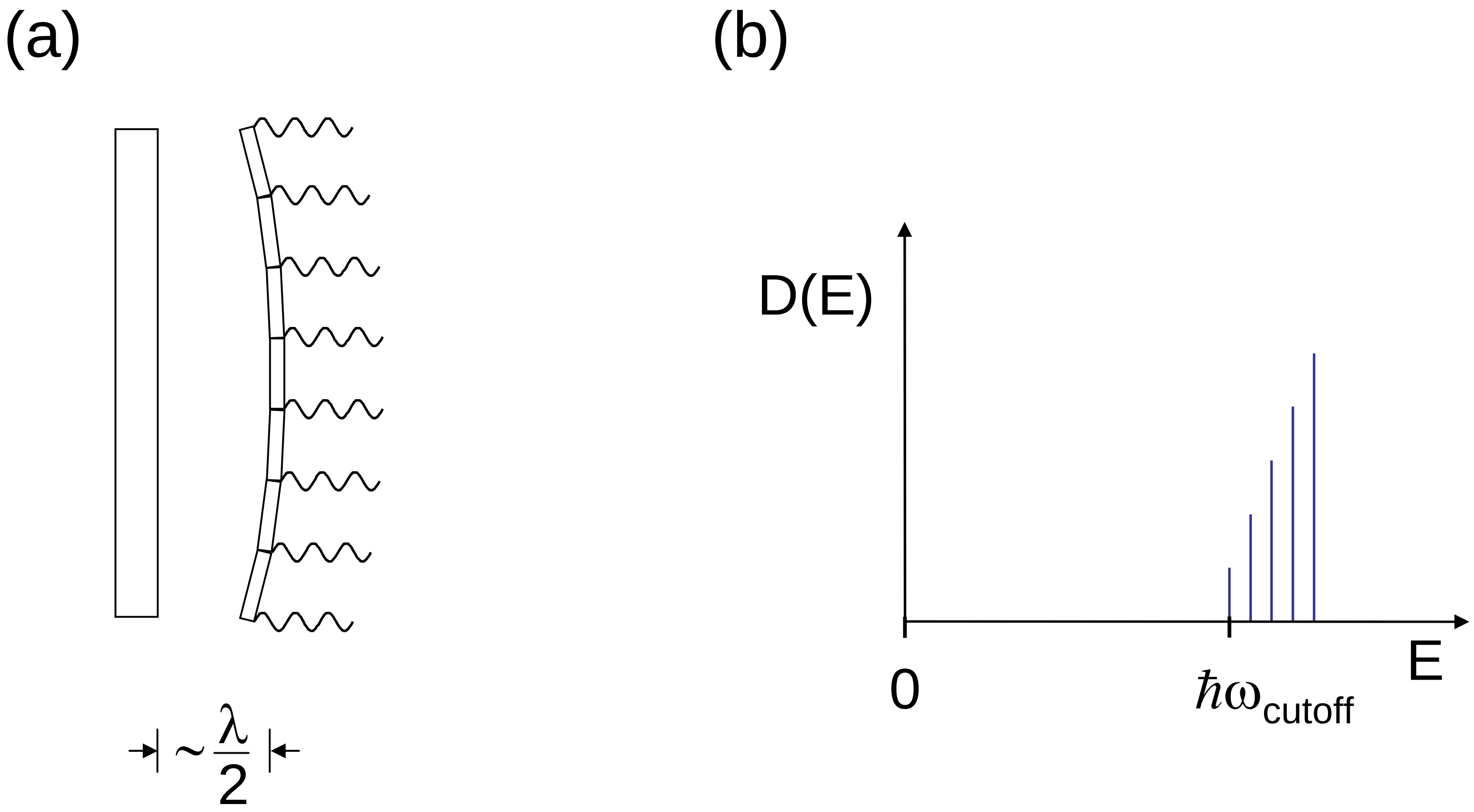

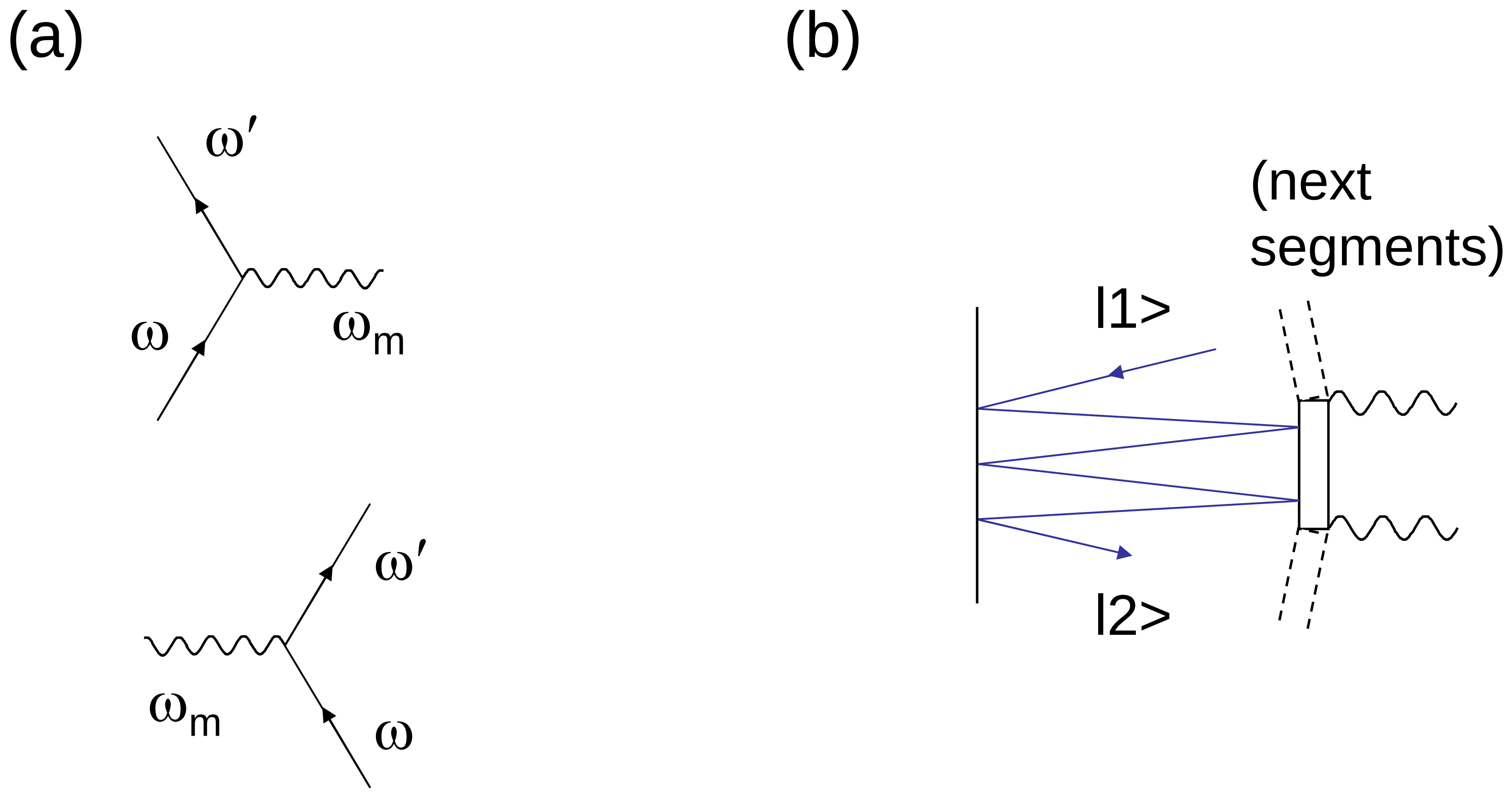

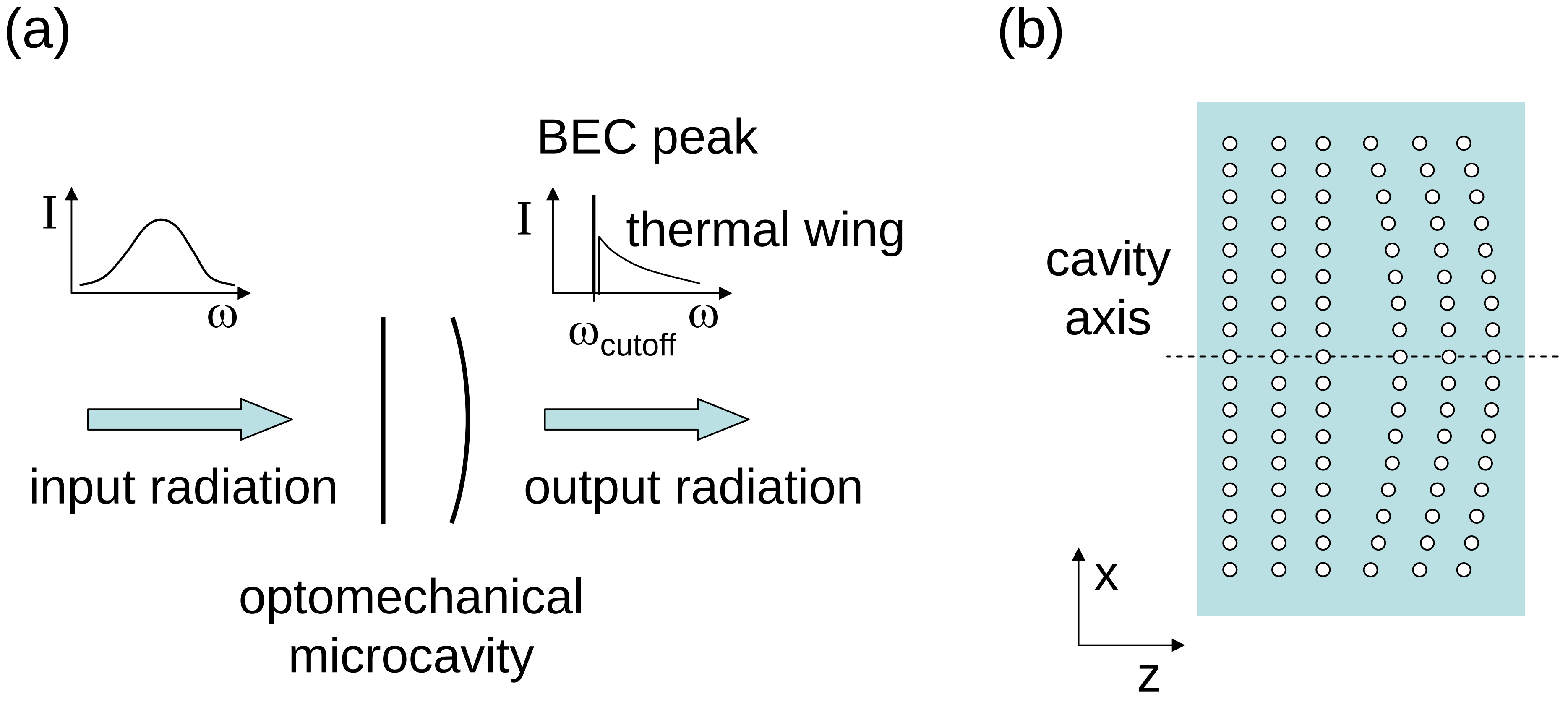